# An Experiment of Research-Oriented Teaching/Learning

Dexin Lu[1,2]*, Dong Ruan[2], Wang Xu[1], Nianle Wu[2,3], Minwen Xiao[4], Yu An[2]

[1]Department for Intensive Instruction, Nanjing University, 210093 Nanjing, China

[2]Department of Physics, Tsinghua University, 100084, Beijing, China

[3]Center of Advanced Study, Tsinghua University, 100084, Beijing, China

[4]Department of Physics, Nanjing University, 210093, Nanjing, China

We introduce our experiment of research-oriented teaching mainly in Nanjing University and Tsinghua University, China. The great population and enrollment in China makes it worth to concern. It lasts 20 years and involves thousands of students and hundreds of instructors, consultant experts. We tried many characteristic styles such as integrated teaching and case analysis, open resources, interactive mode, course paper program, elite solutions and so on. The research on the contents is also placed on the agenda. Many students joined research works that lead to PRL, APL, Nature, Science, and Cell papers. To impart colleagues the essence we offered some examples in every session. We declare the accomplishment of the experiment through this paper and new project is programming.

## I. Introduction

New turn of education reform in China began after "Open to the world" Policy since 1980s. In research universities people started to show interest in research-based teaching/learning or inquiry-based teaching/learning. Our experiment of research-oriented teaching/learning starts about 20 years ago, first in Nanjing University (NU). Later it is extended to Tsinghua University (TH), Zhongshan University, Lanzhou University, and etc. We conduct it mainly through course of Introductory Physics. Facing the great enrollment in China, such an experiment is definitely meaningful. The largest lecture, including 9 classes, contains 384 students. Our earlier practice is recorded in a 1990 paper[1]. In the paper we can find the basic ideas and main frame of the experiment, for example discussion about the relation between knowledge and comprehensive capability, basic research training, course paper, visualized thinking versus logic thinking and advanced visualized thinking, interactive dialogue etc. In the paper we mentioned that a collection of course paper from class 87 was compiled. A student of that class went to Canada later, bringing him with his course paper about Bose-Einstein Condensation. He entered McMaster University and got favorite arrangement. Even earlier event was about "precipitation". Young meteorologist Professor Z Tan now in charge of teaching affair in Nanjing University recalled that instructor Lu went to their dormitory and discussed with him about his research topic of atmosphere precipitation in his sophomore period. In 1980s China it was rare.

Our experiment is quite open. After 10 years practice, in about the middle of 1990s, we were known by more and more colleagues and many of them found it interesting, encouraging and had gradually joined. Meanwhile we began to introduce our experiences to more colleagues and exchange our ideas with colleagues on many occasions. Mr. Lu was invited to give presentation different institutions. From the end of 2003 to next summer vacation, he visited many universities in different cities, flying about 20 269 km. We have also held special workshops in NU 1999.1 & 2004.8,

TH 2003.12, 2006.6, pushing the reform experiment forward[2-6]. Besides the peer instructors in physics fields, the experiment has been concerned by varied communities. Then we may list as colleagues in different disciplinary, researchers of pedagogics[7], and senior researchers, and so on.

Authors all have their own research[8-14] and teaching, covering theoretical, applied and basic physics and mathematics or computational programming. In this paper we only describe our "part time job"——joining the experiment headed by Prof. Lu, mainly in introductory physics courses, and offer some experiences and share with colleagues.

Paper is arranged as follows. Section II is about our main ideas. In Section III we enumerate our main measurements in our research-oriented teaching: Integrated teaching and case analysis; Open resources; Interactive mode; Course paper; Solutions beyond normal; Research and update of contents. Conclusion is drawn in section IV. In Appendix we list some titles of students' paper.

## II. About Research-Oriented Teaching/Learning

In China's community of education many people talk about research-based learning, inquiry-based learning. Everybody has his or her own understanding about it. It is almost definite that many of them are influenced by "Boyer's Report"[15]. In that report the first of ten ways to change undergraduate education is "make research-based learning the standard". Besides that report we also investigated topics from earlier documents[16-18]. That is a part of the background investigation in our experiment. Actually we have examined the situation of China's education in 1950s, 1930s, even ancient Chinese tradition about education and teacher. We have considered how to organize the education, to arrange the curricula, and to design the courses, for some of us are/were head of school or department. The research-oriented teaching/learning mode affects all those aspects substantially.

The following are two paragraphs from reference 15:

> *Undergraduate education in research universities requires renewed emphasis on a point strongly made by John Dewey* (1859-1952) *almost a century ago: learning is based on discovery guided by mentoring rather than on the transmission of information.*
>
> *The non-researcher is too often limited to transmitting knowledge generated by others, but the scholar-teacher moves from a base of original inquiry. In a research university, students should be taught by those who discover, create, and apply, as well as transmit, insights about subjects in which the teacher is expert.*

We think such guiding or mentoring, is a kind of research-oriented teaching. Now we describe our understanding about research-oriented teaching/learning in the following. Firstly we talk about scientific research. Research requires spirit of science, knowledge, ethics, scientific thinking, science literacy, insight, judgment, critical thinking, collaboration, devotement … etc. Our experiment of research-based teaching tries to incorporate those elements in and improves the quality of teaching/learning. It is a sort of style: open, interactive. We encourage discovery and creativity, deliver ideas from researchers. Teaching is not simply transmitting of knowledge and information. Instead *we take knowledge and information as carriers to deliver ideas from researchers*, mainly ideas about science and research. An ancient Chinese proverb reads "Give a man a fish, and you feed him for a day, teach him to fish, and you feed him for a lifetime." It is often cited as an important idea about pedagogy and believed from Chinese Taoist Philosopher Lao Tzu (c. 600 B.C.E). We have not found the proof yet and are not sure about the origin. We accept in that era it is an advanced idea: not just transmitting knowledge or information but putting emphasis on fostering ability. Nowadays somebody still consider it ultimate goal of teaching. However, we think it's insufficient for modern science education. Talking in a way of "fish-fishing", we can discuss with students many ideas such as related ecology, gene etc.

As to the process of teaching, *we implant research training*. Why do we need research training? It can strengthen students' enthusiasm of discovery and creativity, broaden their coverage of knowledge. It may provide strong background for future research: knowledge, consciousness of creation, skills etc. On the other side, it deepens their understanding of the fundamental principles and the essence of physics. So it fulfills the normal requirements of introductory courses well and improves the quality of undergraduate education well.

We understand that research-oriented teaching/learning is suitable best to research universities. In research universities there is a better atmosphere of research. They have needs for research personnel and can make mature evaluation on those candidates. In such environment research training in teaching is definitely welcome. Confined by our experience, we can not tell about applicability in other universities. But we believe many elements are commonly applicable to all and expect to see experience of practice in other type of schools.

## III Main measurements

We enumerate our main measurements in our research-oriented teaching: **a**. Integrated teaching and case analysis; **b**. Open resources; **c**. Interactive mode; **d**. Course paper; **e**. Solutions beyond normal; **f**. Research and update of contents.

**A. Integrated teaching and case analysis**

Knowledge is often taken as central subject in learning and teaching. In his paper "On Teacher" Han Yu (768-824), a Chinese writer of Tang Dynasty, said: "Learning necessitates an instructor, on whom we depend to impart knowledge to us, teach us the way of life and solve our puzzles."[19, 20]

In IT time or knowledge blast time, it is neither sufficient nor appropriate to keep knowledge as center of teaching. The knowledge an instructor can offer are quite limited. Students can gain great amount of knowledge from internet and many other resources easily, almost instantly. We may integrate some points into cases and express some live idea(s) which students may not have many chances to see them in textbooks or other resources directly. We expect those ideas may be kept in memory and as a part of life experience even when the knowledge is forgotten. To some extent it is similar to the situation of the Quote from Einstein[21]: "Education is what remains after one has forgotten everything one learned in school."

Composing of case depends on instructor's background and experience. It is in the process of teaching instructor fills his (her) own experiences and ideas to the subject. Those experiences and ideas enrich the topic and will attach additional value to the topic. In the following we list three cases (a)-(c) as examples to show how we have done.

(a) Kepler problem[22]. In the class we cite a paragraph from Cal Tech's textbook[23]:

> *The task of deducing Kepler's laws from Newton's laws is called the Kepler Problem. Its solution is one of the crowning achievements on Western thought. It is part of our cultural heritage just as Beethoven's symphonies or Shakespeare's plays or the ceiling of the Sistine Chapel are part of our heritage.*

and meanwhile show students the Stories of Genesis on the ceiling of the Sistine Chapel. Students are interested in the culture aspect.

On the other side we present the problem in a way simulating to the research process instead of equations-standard procedure-solution mode. We implant the discussion about central force-conservation of angular momentum-Kepler's second law, effective potential-orbits and so on. We try to let students know: In research how to collect and handle data (Kepler used data collected by Tycho based on systematically 21 years observation), deal with the relation between mathematics and physics, adjust research procedure, understand the meaning of results, extend fruits if possible and so on. For several time we heard Samuel C. C. Ting's talk. He enumerated the history of discoveries in accelerators that he experienced and gave us his comment. It roughly reads: *It is a universal phenomenon that a series of fruits obtained in research is more important than the original purpose. Initially Columbus only intended to find a new way to India; finally unexpected result is he found New Continent.* We directly show students the passage and discuss the meaning of it.

(b) Specific heat of Solids. In this case a brief introduction about Einstein model and Debye model is surely necessary and interesting: their "rude" assumptions and success. After that instructors simply present a picture including four phonon spectrums: Einstein model, Debye theory, Blackman approximation, and real spectrum. Meanwhile let students comment "Why can so rude assumption lead success?" "Which is creative work? Which is only good work?" In most case half to half students think first two theories are creative and two think the third. It may have underlying influence on their future research career.

(c) Drude's discovery. In 1900 without background of quantum and quantum statistics, it is impossible to obtain correct value of electronic specific heat. But with classical statistics and thereby two mistakes, together with an additional numerical error Drude published perfect value of Lorentz number in Wiedemann-Franz law. Now our topic is how to comment a theory which is perfectly coincident with empirical result. Some students said it is the first time for them to know a "perfect" theory could be based on totally wrong mechanism.

There are many subjects of science education; there are many ways of same order to compose cases to express.

**B. Open resources**

In an introductory course we should avoid to restrict the students in a textbook, often only one textbook. We advocate open course (we do not mean class should be open to more audiences or public instead we mean open to more resources). An open course can make teaching/learning dynamic. It can let students change from learning physics course to learning physics. By open we mean:

*Open to references*

*Open to internet*

*Open to the open problems*

In our textbook more than 200 reference papers 56 books are cited[22]. In the software of supplementary material we recommended a very short paper of 216 words as a start.

To encourage our students we quote "*It is of great advantage to the student of any subject to read the original memoirs on that subject, for science is always most completely assimilated when it is in the nascent state.*" by James Clerk Maxwell. For a freshman it is really a challenge to read even only 10 original papers. However it is of great advantage if a student keeps the reading habit.

Nowadays in China all universities and most middle schools campus network is already very popular and serves as important constituent of education. Platforms based on campus networks offer many related materials. This is helpful to most students, for they can obtain information and knowledge easily and quickly. But most of those materials are organized or compiled by teachers. We encourage students to find the original references. Sometimes we recommend some originals as clues.

Open to the open problem. It may cause controversy if we introduce some open problems in course. But for a high level course it is worth to try. We introduced problems from small as puzzle in assignments to large as dark energy, or the left-hand-material which challenges the common senses. In most cases the solutions of the topics go beyond the scope of the course. But the topics themselves do bring students sense of mission. They feel they are the successors.

It is possibly interesting that we offer a sample of utilizing internet to study. It comes from an "incident". Once a student asked a question in the class: why the standard kilogram should be sent for comparison with the international prototype of the Kilogram timely and how can we make sure which is real standard? The instructor makes a direct Google. 4380 items show the history of the kilogram and the future, far

beyond the material offered in textbook and by instructor. For example, we found the following passages

*A means of defining the kilogram in terms of electric units has been proposed by B. P. Kibble...*[24]

*To update the kilogram, Germany is working with scientists from countries including Australia, Italy and Japan to produce a perfectly round one-kilogram silicon crystal. The idea is that by knowing exactly what atoms are in the crystal, how far apart they are and the size of the ball, the number of atoms in the ball can be calculated. That number then becomes the definition of a kilogram*[25].

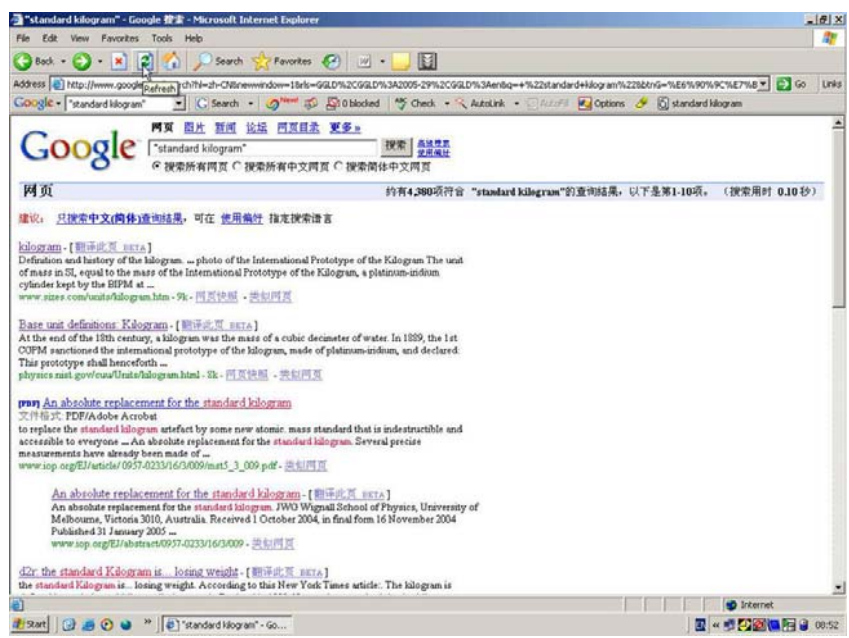

**Fig. 1. Google "Standard Kilogram"**

Since then we take it as a normal episode. And students feel "All are equal before internet; network world is rich and colorful", not only games and chat. We think internet is a very important resource.

Also we often quote materials from internet: science news, e.g. Nobel Prize, important progress, science scandals (as a part of ethics education, see part III-d) and so on.

Moreover we think teacher can serve as an active resource. The status of an instructor as resource implies a statue of students as center of learning. By active we mean who offers not only knowledge and information, but also ideas and suggestion about literacy and capacity. We ask students for comment about difference between an encyclopedia and a professor. Students appreciate the dynamic aspect of the active resources. Here we should mention the prominent American psychologist Carl Rogers. He wrote[26] "I see the facilitation of learning as the aim of education". He saw facilitator as…and serving as a flexible resource[27]. We find some similarity between the viewpoints about resources. Carl Rogers asserted that the student has interests and enthusiasms, and the task of the teacher was to free and to aid these interests and enthusiasms[28]. We find another similarity here. We try same thing and also free students' potential.

**C. Interactive mode**

Interactive mode provides *reciprocity*: faculty can learn from students as students are learning from faculty. It improves the quality of teaching/learning substantially. Adoption of interactive mode also means the equal status of teacher-student. It requires instructors realize that students are the center of learning. However teacher should do more. It is instructors' obligation to create an atmosphere of free communication and offer more chance to discovery.

There are many versions of interactive mode (a)-(e).

(a) ***Passive mode***: When it becomes one of standards for instruction quality evaluation, the instructor may feel some pressure. In some classes the instructors ask frequently

"Is it so?" "Right?" to let inspectors (experts) know their lectures are interactive. It is not really interactive; we may call it *passive-mode*.

(b) **Vote mode**: A simple mode is *vote-mode*, teacher asks students to answer "yes" or "no". It allows students limited time to prepare and some students simply guess without think thoroughly.

(c) **Statistical mode**: A better one we would like to call it *statistical-mode*. Professor Richard McCray (JILA, University of Colorado, CO, USA) designed sets of questions and let students choose the answer. Clickers (a kind of voter) are used to feedback the choices and receiver shows the statistical results instantly. Then he can easily judge what students need or how to correct unexpected tendency. This version is especially effective for large classes. It can accurately locate problem and help to control the progress of lecture.

(d) **Induce mode**: Another one we call it *induce-mode*. Professors Art Hobson (University of Arkansas, AR, USA), designed sets of questions let students join. For example a question about energy states of two magnets, where energy exists etc induces students to accept:

    *"Empty" space can have mass.*

    *Further more—*

We attended the workshops that demonstrate latter two modes as "student" and other classes as looker-on. Our feeling is that all modes are helpful to students' participation and to different extent lead to the results the instructor expected in advance. The latter two modes or similar seem to be quite efficient and are worth to be recommended.

(e) *Student initiative mode*: We have some different stories. We emphasize the inquiry in class and welcome or concern unknown results. Students are free to speak out in the class and express their ideas without pressure. In the following we list two events (a), (b).

> (a) *Problem 2.16*[22,29]. A virus particle of mass $m$ in solution in a centrifuge is, at a particular moment, at distance $r$ from the axis of rotation and moving radially outward at a relatively constant speed $v_0$. The centrifuge is rotating at $\nu$ rev/min. Discuss the motion quantitatively, giving the magnitude of all forces and accelerations as viewed from a reference frame (a) rotating with the centrifuge and (b) fixed in the laboratory.

This is a "mature" problem. Once in a class a student raised "How about the motion other than that particular moment?" He felt "constant speed is not possible again" and oppugned "Does that particular moment always exist?" His question might be regarded as nonsense and neglected. However, in an "interactive class", it is a fresh idea that interests the instructor. Actually the instructor had no idea to provide a quick solution/answer at that moment and felt it was a challenging question. He initiated the discussion with students. Simple-minded conclusion is accepted: "In case speed is constant, the resistant from liquid is constant no matter what power of speed depends on and inertial centrifugal force is proportional to $\omega^2 r$. It is impossible those two forces cancel except at certain position (corresponding that particular moment)." As matter of fact, this problem is more complicated than that. The density of virus could be a parameter, e.g. anthrax virus without water. Moreover the inertial centrifugal force may lead to effective buoyancy. The virus can be ever-floating, ever-sinking, oscillating or critical.

(b) ***Problem 12.8***[22,30]. A bullet at a temperature of 60°C and traveling at a speed of 400m/s strikes and becomes embedded in a large barrier at 15°C. The bullet has a mass of $20 \times 10^{-3}$ kg and a specific heat of 400 J/kg·K. What is the total change of entropy? (2003.12.19)

It happened on Dec 19 2003 at Room 6c101, TH. It is a demo-lecture in a workshop on research-oriented teaching, with 70 professors from nation-wide universities there. After some routine discussion, a student suddenly oppugned: "Why don't you discuss over the contribution from the shot-hole in the barrier?" All laughed, since the question looks quite funny and worthless. To all's surprise, the instructor said something impressive "Well Mr. Yan raised a problem and I am afraid all teachers of thermal physics over the world have never considered. Making a shot-hole is irreversible and the entropy increases. I do not know how to calculate it now; perhaps it can be calculated as configuration entropy."

The examples above show that besides the free atmosphere the instructor's ability of identifying the new problem is decisive. This kind of interaction is considerably valuable but is not easy to handle. Such a situation often makes instructor uneasy. Every instructor has the right to choose and design the suitable mode.

Discussion session is important peer interactive activity. Instructor's task is to search and design problem that students can discuss and inquire by knowledge they already have. We encourage students challenge existing conclusion, propose their new scheme and opinions. Sometimes important thing is not conclusion but the process and feeling they experience.

Other interactions appear through email and in presentation session. We will show them in next section.

**D. Course paper**

Course paper program in the introductory courses is feasible and works well. It may provide strong background for future research: knowledge, consciousness of creation, skill training etc. To July 2005 the total numbers of course papers in our program are 4682 (NU and TH). In 2005-2006 academic year we have 820 more papers (NU and TH).

In our program, we require course paper exactly like a normal academic paper: from the topic, the content, layout, reference citation etc. Publication is not our goal but standard. Moreover we incorporate in ethics education; discussions about pseudoscience and science scandals, superstition, pathological science[31], amateur scientists and so on.

Running the program we do three related things: supplement of materials in lectures; group mentoring; and one-to-one tutorship.

In lectures we have done the following as parts of normal:
> *to point out the innovation of an important fact;*
> *to pay more attention to the events of discovery;*
> *to hint the clues for imagination and development;*
> *to emphasize scientific ethics;*
> *to enhance scientific literacy training and so on.*

In group mentoring: we try

*to enumerate the ways to find subjects;*

*to introduce creative topics;*

*to comment on excellent or trivial topics;*

*to show examples of references for topics confirmation and legal citation.*

Many colleagues have a misery that they could not create/offer students topics. They feel it's really difficult to find some topics which are both new and appropriate for college students. We did not offer topics to students. Instead we show students the way to find topics. In the following we show an example about how to find topics. Starting from simple pendulum, we let students describe what a simple pendulum is. It is not so difficult to list at least the following (approximation or simplification) conditions as shown in the figure.

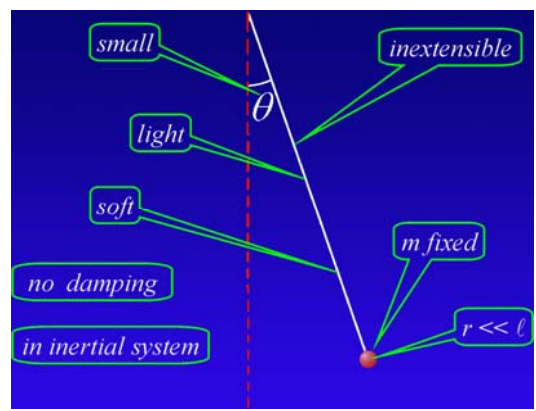

**Fig. 2. Simple Pendulum**

Then we offer an idea: "remove or change any one or some condition(s) we will get a real problem". Totally we will have:

$$C_8^1 + C_8^2 + C_8^3 + C_8^4 + C_8^5 + C_8^6 + C_8^7 + C_8^8 = 2^8 - C_8^0 = 255$$

Surprised? One step beyond the simplest example will provide us hundreds of topics. Indeed we can obtain much more solutions from different initial conditions, motions in two perpendicular planes, charged bob etc. But they are trivial. A student considered a pendulum with charged bob in a magnetic field and found the Lorentz force is of same mathematical structure as Coriolis force. Then the electromagnetic pendulum is similar to a Foucault pendulum. It is a quite meaningful discovery and breaks our conclusion about trivial topics.

We also list other ways about topic finding: reverse (e.g. inverse Compton Effect); change of dimensionality (scale); combination; transplant (fraction-friction); from experiments; from hobbies and games (kite, soap bubble, stone-skipping, swing, top); from topics done and so on. A normal procedure is recommended to students: start from a phenomenon, find its kernel physics, extract key words, search and browse, find references, and repeat. Some students take many cycles to determinate their topics. The training is tough but fruitful. It is emphasized that research refuses replica. A topic can be done only if it is supported and confirmed by references.

One to one tutorship is necessary, for every student has his (her) own topic and different problems: the correctness of the topic; degree of difficulty of the problem; references, and technical problems. Such individual tutorship proceeds mainly through email or studio. We promise to reply within 24hrs.

Now the larger classes will bring teachers heavy burden. In rush period, when you open your mail box you might find 100 mails from students who want to discuss their course paper topics with you. Students are happy to have chance to converse directly. But for teacher as supervisor it is really a hard time. Besides the responsibility and passion supervisor should have strong research background and wide coverage of knowledge. While an experienced instructor can treat most of the mails in shorter time and concentrate to 2 or 3 ones that take time to tackle, a newer can be trapped in mails. It is this part of the course some colleagues feel fearsome. Actually accumulation of experience is quite important. It's observed that an instructor often feels much easier to handle the case in his second turn.

We adopt three schemes to solve the problem.

Firstly we construct a database of course papers, it is finished in TH spring 2006. The database provides many original materials: references, topics, and even some technical calculation or plotting. It is imaginable the database also offers hints to lecturing and group monitoring. For example we extracted some subjects as example of text.

In NU we are supported by a "map of experts". It starts to form at a quite earlier time when we faced students of science class who are physics, astronomy, chemistry, biology, geology, geography, and mathematics majors. The consultation and evaluation on topics involving inter-disciplines are difficult. The experts in corresponding fields may help a lot. They are so professional and can give an accurate comment in a minute. They often feel the problem interesting and realize our reform program meaningful. We appealed them often and encouraged students to communicate with them directly. Indeed quite possibly they are helping candidates of their future graduates.

Thirdly we start a new improvement in NU spring 2007: in first semester of the course students complete a topic proposal instead of a paper; and in the following semester they finish the paper. Now this scheme is tested in TH. Under this scheme both teacher and student sides feel easy. Students have more time to consider and carefully find the references they want. From reference-finding alone we feel students are really doing what they enjoy. In June 2007 we praised some students' efforts. The class reaches to a new level at least in four aspects. 1. A student finds the earliest references ever appeared in our project: they are three *Nature* paper about golf[32-34] since 1890 (we mentioned old subject needs old and frontier references). 2. A student lists references more than a hundred. 3. A student finds three PRL papers of falling paper[35-37](it is not easy to locate them according to normal procedure). 4. Discussion involves great diversity (Mpemba effect, Brazil nut effect, broken vase theory, Einstein tea leaves paradox, silo effect, freak wave, creeping of earthworm etc).

Constantly renewal of the course paper evaluation system is also helpful. In 2006 a renewed operation procedure showed much higher efficiency. At first five TAs and three instructors took two weeks to complete the evaluation of 256 papers. Using the new system, in next semester we took three days finished the same work. TA group and instructor group offered quite similar scores of as 29.2% as high coincidence. New system also includes an appeal item. Invoking this item a student required re-evaluation. We invited two professors to do it. The result was not different from the original but the student felt that an academic paper author's right was respected. For plagiarizer the new regulation gives severe score deduction as a caution, the policy remaining effective within four years if we can decide the fact of plagiarism at later time. We processed four events for that class.

What has academic level of the papers reached? What kind topics have our students chosen? Quantitative description is almost nonsense. We attached an appendix to list some topics our students have done.

**E. Solutions beyond normal**

In China's middle schools besides the exercises after every lesson teachers usually collect or compile many others and organize intensive trainings as a tactics. Many students can complete exercises very quickly, nearly like conditional reflection. After entering university, while some students keep the tactics as habits, others doubt its necessity. In such a situation how should we organize our introductory courses? Even though the chain of question- exercise- problem is basic, simply repetition of middle school mode will bore students.

We adopt "less but better" strategy, i.e., reduce the amount of the assignments and require the better quality. By the better quality we mean the recommendation of high level techniques and the introduction the charm and beauty of the physics itself. Further requirement is to build up spirits of criticism and inquiry.

We remind students use their "legal right": when solving problem they can query the rationality of conditions in the problem; they are "legal" to give comments on quality of problem; they have the rights to revise the problem, make it better or extend the result if possible etc. We tell students they can probe into various possibilities of solving, and do the exercise as a project.

For many years we have accumulated some excellent demo-examples of problem-solving. Those solutions don't involve a lot of math-difficulties and/or artificial compositions, but clear physical ideas and deep inquiry. In the following we list three of them (a)-(c). Problems are normal, but solutions are unusual, that many colleagues called as elite solutions.

(a) **Problem 3.4**[22,38]. A very small cube of mass $m$ is placed on the inside of a funnel rotating about a vertical axis at a constant rate of $f$. The wall of the funnel makes an angle $\theta$ with the horizontal. If the coefficient of static friction between surfaces is $\mu$. What are the largest and the smallest values of $f$ for which the cube will not move with respect to the funnel?

It is straightforward to obtain a lower and an upper limit of rotating rates. If we consider the concept of the critical angle and take the coefficient of static friction and angle $\theta$ as parameters, we get the following "phase diagram".

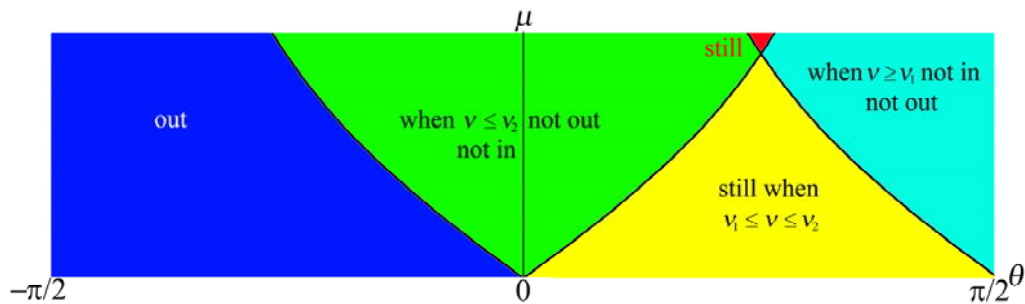

**Fig. 3. Phase Diagram of Rotating Funnel**

Students are excited to see the diagram and some students suggested 3D diagram is possible if we take $f$ as the third parameter.

(b) **Problem 6.3**[22,39] A length $\ell$ of flexible tape is tightly wound. It is then allowed to unwind as it rolls down a steep incline that makes an angle $\theta$ with the horizontal, the upper end of the tape being tacked down. Show that the tape unwinds completely in a time $T = (3\ell/g \sin \theta)^{1/2}$

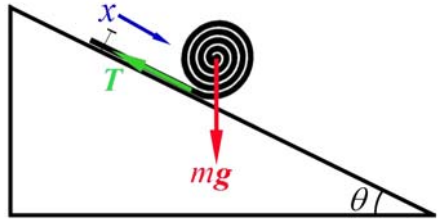

**Fig. 4. Unwinding Tape**

Since 1977 students (and teachers) have been doing it until someday 2005 in 6c300 TH, a student of class 2005 objected to the rigid disc approximation: "Anyway the radius keeps deminishing!" The instructor thought his statement is reasonable. With several months' effort we found beside geometry relation, dynamic effect and variable mass effect should also be considered. The content has submitted to "Physics and Engineering" and published[40]. At that time we did not realize that it has a profound background. Here we list some references we found afterwards and would like to apologize to readers for careless omission of so important references.

Early in 1941 L. L. Pockman[41] has drawn attention to this problem with a note entitled "Nonconservation of energy—a paradox": the gravitational potential energy of the tape after unwinding completely will be less than the initial potential energy. It is the problem we feel puzzled first. I. M. Freeman[42] considered the situation of tape on a horizontal plane and, according to energy conservation, calculated the squared velocity. He already mentioned the relation we listed as Equation (5). Freeman considered the case of a thin tape wound on a massless cylindrical bobbin, so that the radius remains constant and calculated the complete time of unwinding. Relative to our later solution two situations are oversimplified, it is first dynamic study. Calivini[43] discussed thick tape, i.e. expansion of a spiral (we only mentioned this point).

(c) **Problem 3.9**[22] At 30°05′ N latitude, a body falls from 10m high. Find the horizontal deviation due to Coriolis force (air friction ignored).

Another example is **Falling-body paradox** or **Coriolis puzzle**: falling-body deviates to the east due to Coriolis force. It is easy to integrate the equation of motion. Occasionally students do it as horizontal projectile problem for Coriolis force is substantially the kinematical effect:

$$\Delta v \cdot t = [\omega(R_\oplus + h) - \omega R_\oplus] \cdot t$$
$$= \omega R_\oplus t$$

The physics picture of this scheme looks so clear and the calculation is so simple. Moreover we propose one more scheme: do it by the elliptical trajectory. Three schemes offered different results with total different numerical coefficients, constructing a paradox. Why is it so? How to reach correct understanding and solution? Now as an example we show why the scheme 2 is wrong. In scheme 2 the flat approximation sounds good: the deviation is so small, for a height of 10 m, it is only order of mm. But it is misunderstanding. Both the projectile and the plum point move distances over $10^2$ to $10^3$ m, the small deviation is the difference between two large distances. The accuracy is unjustified. The curvature of the surface and nonparallel gravity make the solution simply wrong.

Such examples really stimulate students (even instructors) to dig deeply and explore the physics in it. May 2006 when the tape-unwinding was talked in HUST (Huazhong University of Science and Technology) students soon proposed more than 20 schemes for further investigation, including computer simulation, experimental investigation, and so on. "Never see students so zealous to a single problem." some instructors said.

Discovery of such solutions beyond normal is quite difficult. It's even harder to decide which problem is suitable candidate. Often a good finding takes long time, say two or three years. Colleagues call on publication. We will do it and wish their charm arouse more colleagues' enthusiasm to discover those elite solutions and exchange them.

**F. Research and update of contents**

Research-oriented teaching is a paradigm. It is not simply a method of teaching: open, interactive etc. Besides 'how to teach' we have to decide 'what to teach?' i.e., decide and offer a characteristic syllabus. Moreover it should include research on the contents of teaching. This might lead to revision, renewal and update[44-46].

We have carefully considered the expressions of some topics[22, 46]. We showed different opinions to students and told them the possibility of different understandings. It is helpful to realize how academic criticism goes along and how to pursue better or correct physics. In the following we list four cases (a)-(d). (a) Is traditional kinematics complete? It is a kind local description. If we incorporate trajectory in phase space in; it offers global description. The phase space is not new and usually is discussed in statistical physics. In kinematics part a discussion about phase space is meaningful. (§2.5 in reference 22) (b) Are thermodynamic potentials extensive? Indeed it is conditional. For example the thermodynamic potentials of a self-gravitational system are not extensive. If such expressions appear in monograph or research paper why should not they appear in textbook? (§12.4 and 12.6) (c) In discussion of sp hybridization we found the nodal surface deviates from the origin due to a simple linear superposition. This conflicts to the diagrams in chemistry textbooks and is perplexing. Indeed the location of nodal at the origin is a result of special linear superposition, so-called Slater AO (§26.1). (d) In discussion f Hubble's law a reasoning of "the redshift-distance law − recessional redshift − the velocity-distance law" is often used and is questionable (§9.2). Redshift of cosmological origin includes Doppler Effect and the influence of background gravitation. Hubble's law proves cosmological principle (§32.1). Robertson-Walker metric is used to discuss cosmological redshift (§32.4). It is at the right time to write this paper we find a professional reference[47]. It is so helpful to us for the deeper understanding of the topic and further improvement of expression. The coverage area of an introductory course is in general much wider than instructor's research field, the related professional papers are necessary for instructors to keep the content scientific and veracious. For a historic topic we had a wonderful encounter, a similar situation. That is about Stern-Gerlach experiment. In some textbooks the space quantization was immediately attributed to spin. We wonder how Stern-Gerlach can explain their experimental result in 1922 with a 1925 idea, spin, from Uhlenbeck and Goudsmit. It is not so clear until we found a paper[48] by Friedrich and Herschbach. "The earliest attribution of the Stern- Gerlach splitting to spin that we have found did not appear until 1927, when Ronald Fraser noted that the ground-state orbital angular momentum and associated magnetic moments of silver, hydrogen, and sodium are zero"[48]. The history of Stern-Gerlach experiment is indeed very instructive. We believe and adopt the story, rewriting the paragraph.

Responsible authors of textbooks usually keep a good renew rate, keeping the readers aware of new progress in the related fields. Duteous instructors should try to introduce the latest developments to students timely. Such interpolation could make course dynamic, and enhance students' sense of mission. In the following we list some of those new contents (a)-(d). (a) In chapter 4 we introduced new resolution by IAUGA (§4.1):

> On 24 August 2006 the International Astronomical Union General Assembly resolves: Pluto is a "dwarf planet" by the new definition and is recognized as the prototype of a new category of trans-Neptunian objects.

(b) No doubt the dark energy is one of the most noticeable new concepts. We mentioned it (§32.5) and linked it to Einstein's blunder (§32.4), placing a cosmological constant in his field equation. (c) "Solar neutrino problem" and neutrino oscillations are introduced in §31.1 when we introduced 2002 Nobel Prize. (d) 2006 Nobel Prize in Physics is awarded to Mather and Smoot for their discovery of the blackbody form and anisotropy of the cosmic microwave background radiation as well as its small variations in different directions. It has been mentioned with 1978 Nobel Prize to Penzias and Wilson (§32.1).

Finally we also mentioned some controversy. Excess heat[49,] (§29.5) was considered as pathological science. Later it was reported[50,51] that the experiments show higher replicability. It is controversial to introduce controversies; but we can not evade them for we do not live in vacuum.

## IV. Conclusion

It is about 20 years passed since we started our experiment on research-oriented teaching. We would like to declare the accomplishment of it through this paper. Evaluation of it can be left to our colleagues or even posterity. What we concern best is if it attracts students. It is a necessary precondition and absolutely important to the raise or improvement of the quality of our teaching[52]. Heads of our school or department never worry about recruitment of students and always please to know so many graduates still stay in the fields of pure sciences and join research. Mr. J Wang of Class 96 published a paper[53] on Nature Structural Biology during his graduate study. His advisor comments: "One important reason that he can publish such a paper is that he studied general physics well". Many students of ours published their papers during their graduate periods (for example, Science papers by Junjun Wu[54] of class 95; Congjun Wang[55] of class 99; Xiaoshan Xu, Shuangye Yin, and Xuemei Cheng[56-8] of class 97; Congjun Wang). In 2005 we have an influential Nature paper[59] (Jun Lu of class 97) and three Cell papers[60-62] (Fei Sun, Hui Jiang, and Ji-Song Guan of class 2001). Students often said that their research consciousness initiated in those introductory courses. They affirm that training in those courses is definitely fundamental to present or future research career.

## Acknowledgement

We would like to thank Professor Shiqun Li for instructive discussions, thank Professor Y Wan for locating reference 19.

## Appendix: Students' paper title

| 1 | SHI Leiming | On effect of gravitational retardation on motion of stars |
| 2 | WANG Jun | Visual appearance of moving objects with high speed |
| 3 | CHENG Xuemei | Explanation of the relation between static friction and normal, contact area by fractal theory |
| 4 | YANG Chengyong | Comet collides Jupiter |
| 5 | LU Lu | Quantum Well and Quantum Devices |
| 6 | YU Hao | Design of measurement of $g$ by Laser confinement of atoms and its geophysical significance |
| 7 | WANG Junfeng | Reverse Compton Scattering and ICS Model of Radiative Pulsar |
| 8 | KONG Dong | Changeable Demons—Computer Imitation of Fractals in Biology |
| 9 | CHEN Ming | 7-Fold Symmetry |
| 10 | ZHOU Zhe | The Analysis of the Process of the World Trade Center's collapse in the |

| 48 | HASAN GOKCE | Entropy and the Universe |
|----|-------------|--------------------------|
| 49 | HUANG Wujie | A Simple Analysis of Giant Magnetoresistance |
| 50 | YAN Feng | ON Formalism and Initial Entangling of Quantum Games |
| 51 | ZHANG Meiying | A Brief Study of Inverse Compton Effect |
| 52 | GE Xiao | Initial Research of Electromagnetic Foucault Pendulum |
| 53 | CHEN Yibiao | One-dimensional Finite Periodic Potential Well |
| 54 | TANG Liyang | New Model of NN with Consciousness" |
| 55 | GENG Yifeng | The Origin of Triboelectricity |

---

*Electronic address: dxlu@nju.edu.cn

[1] D Lu, Instruction of Introductory Courses and Development of Students' Creativity, Research and Exploration in Higher Education (in Chinese) **6**, 15 (1990).

[2] D Lu, Research-based teaching of University Physics, Physics and Engineering (in Chinese) **14** (1), 1 (2004).

[3] D Lu, Research-based teaching of University Physics II, Physics and Engineering (in Chinese) **14** (2), 1 (2004).

[4] D Lu, Further Discussion about Research-based Teaching/Learning, China Higher Education (in Chinese) **21**, 24 (2004).

[5] D Lu and W Xu, Course Construction Aimed to Research-based Teaching/Learning, China University Teaching (in Chinese) **5**, 20 (2004).

[6] D Lu, On Research-based teaching. Physics and Engineering (in Chinese) **15**, 1 (2005).

[7] F Zhang, Comprehend Dexin Lu's Research-oriented Teaching/Learning, China University Teaching (in Chinese) **3**, 41(2007).

[8] D Lu and K Golden, Third-frequency-moment sum rule for electronic multilayers, Phys. Rev. **E61**, 926 (2000).

[9] Dong Ruan and Weicheng Huang, General realization of N=4 supersymmetric quantum mechanics and its applications, J. Math. Phys. **44**(7), 2787-2805 (2003).

[10] W Xu, and Z Z Li, The slave boson mean-field theory of two-conduction-band Kondo alloys, J. Phys. **C21**, 4083 (1988).

[11] W Xu, and Z Z Li, The electrical resistivity of the two-conduction-band heavy-fermion alloy system, J. Phys. Condens. Matter **2**, 109 (1990).

[12] Z Ma, N Wu and etc. A monolithic $Nd:YVO_4$ slab oscillator-amplifier. Opt. Lett. **32**, 1262 (2007).

[13] M Xiao, Extended ensemble theory, spontaneous symmetry breaking, and phase transitions, J. Stat. Mech., P09007 (2006).

[14] Y An, Mechanism of single-bubble sonoluminescence, Phys. Rev. **E74**, 026304 (2006).

[15] Boyer Commission on Educating Undergraduates in the Research University, Reinventing Undergraduate Education: A Blueprint for America's Research Universities, 1998.

[16] Boyer Commission on Educating Undergraduates in the Research University, S. S. Kenny (chair). Reinventing Undergraduate Education: Three Years after the Boyer Report, State University of New York–Stony Brook, 2002.

[17] Advisory Committee to the Directorate for Education and Human Resources, NSF. SHAPING THE FUTURE, New Expectations for Undergraduate Education in Science, Mathematics, Engineering, and Technology, 1996.5.

[18] John S. Rigden, Donald F. Holcomb and Rosanne DiStrefano, The Introductory University Physics Project, Physics Today April **46,** 32-37 (1993).

[19] HAN Changli Collectanea, Collation and Annotation. Shanghai: Shanghai Ancient Book Press, 1986,

[20] English translation from WANG Wuming. http://www.zftrans.com/bbs/

[21] http://www.quoteworld.org/ Quote from Einstein.

[22] Dexin Lu. University Physics, CHEP Beijing and Springer-Verlag Berlin Heidelberg, 1999.9.

[23] Richard P. Olenick, Tom M. Apostol, David L. Goodstein. The Mechanical Universe: introduction to mechanics and heat. Cambridge, New York: Cambridge University Press, 1985.

[24] Terry J. Quinn, The kilogram: The present state of our knowledge, IEEE Transactions on Instrumentation and Measurement **40**, 81-85 (April 1991).

[25] diego's weblog: the standard Kilogram is... losing weight.

[26] Carl Rogers, Freedom to learn for the 80s, Columbus, OH, Charles E. Merrill, 1983.

[27] Carl Rogers, Client-centred therapy: its current practice, implications, and theory, Boston, MA, Houghton Mifflin, 1951.

[28] UNESCO: International Bureau of Education, *Carl Rogers*. Prospects vol. XXIV, no. 3/4, 1994, p. 411-22.

[29] Textbook the problem we adopted from. (to be found.)

[30] Textbook the problem we adopted from. (to be found.)

[31] I. Langmuir, Pathological Science, Physics Today, **10**, 36 (1989).

[32] Tait, P.G., Some points in the physics of golf. Part I, Nature **42**, 23-420 (1890).

[33] Tait, P.G., Some points in the physics of golf. Part II, Nature **44**, 98-487 (1891).

[34] Tait, P.G., Some points in the physics of golf. Part III, Nature **48**, 5-202 (1893).

[35] Z. Jane Wang and Umberto Pesavento, Falling Paper: Navier-Stokes Solutions, Model of Fluid Forces, and Center of Mass Elevation, Phys. Rev. Lett. **93**, 144501-1 (2004).

[36] Andrew Belmonte, Hagai Eisenberg, and Elisha Moses, From Flutter to Tumble: Inertial Drag and Froude Similarity in Falling Paper, Phys. Rev. Lett. **81**, 345 (1988).

[37] Y. Tanabe and K. Kaneko, Behavior of a falling paper, Phys. Rev. Lett. **73**, 1372 (1994).

[38] Resnick R, Halliday D, Krane K, Physics 5th ed. John Wiley & Sons, 2002. Problem 5-18

[39] Resnick R, Halliday D, Physics 3rd ed. John Wiley & Sons, 1977. Problem 12-41 (withdrawn in later edition).

[40] D Lu, An example of problem-solving in research-based teaching, Physics and Engineering, **16**, 7 (2006).

[41] L. L. Pockman, Am. J. Phys. **9**, 51 (1941).

[42] I. M. Freeman, The Dynamics of a Roll of Tape, Am. J. Phys. **14**(2), 124-126 (1946).

[43] P. Calivini, Dynamics of a tape that unwinds while rolling down an incline, Am. J. Phys. **51**(3), March 1983.

[44] D Lu, Original Intention of University Physics, Research and Exploration in Higher Education (in Chinese) **1/2**, 8 (1999).

[45] D Lu, Design of University Physics Course and Textbook, China University Teaching (in Chinese) **2**, 12 (2000).
[46] D Lu. University Physics, 2nd ed. 3rd print. Higher Education Press Beijing: 2006.
[47] Harrison, E. The redshift-distance and velocity-distance laws, Astrophysical Journal **403**(1), 28-31 (1993).
[48] Bretislav Friedrich and Dudley Herschbach, Stern and Gerlach: How a Bad Cigar Helped Reorient Atomic Physics, Physics Today December 2003, page 53.
[49] HE Jingtang, Arguments on Cold Fusion Research. Progress in Physics **23** (9), 69 (2005).
[50] Peter L. Hagelstein, Michael C. H. McKubre, David J. Nagel, Talbot A. Chubb, and Randall J. Hekman, New Physical Effects in Metal Deuterides, Proceedings of the 11th International Conference on Cold Fusion, Marseilles, France, 31 October - 5 November 2004 edited by Jean-Paul Biberian.
[51] HE Jingtang, Experimental Progress of Cold Fusion, Progress in Physics **25**(2), 221(2005).
[52] D Lu, Ponder about Improvement of the Teaching Quality, China University Teaching **10**, 4 (2003).
[53] **Jun Wang** and Wei Wang (王炜)，Nature Structural Biology **6**(11), 1033-1038 (1999).
[54] Thuc-Quyen Nguyen, **Junjun Wu**, Vinh Doan, Benjamin J. Schwartz,* Sarah H. Tolbert,* Science **288**, 652-6 (2000).
[55] **Congjun Wang**, Moonsub Shim, Philippe Guyot-Sionnest,* Science **291**, 2390 (2001).
Three papers in single issue Science
[56] Ramiro Moro, **Xiaoshan Xu**, **Shuangye Yin**, Walt A. Science **300**, 1265 (2003).
[57] Mingwei Chen,* En Ma, Kevin J. Hemker, Hongwei Sheng, Yinmin Wang, **Xuemei Cheng**, Science **300**, 1275 (2003).
[58] Dong Yu, **Congjun Wang**, Philippe Guyot-Sionnest,* Science **300**, 1275 (2003).
[59] **Jun Lu**\*, Gad Getz\*, Eric A. Miska\*, Ezequiel Alvarez-Saavedra, Justin Lamb, David Peck, Alejandro Sweet-Cordero, Benjamin L. Ebert, Raymond H. Mak, Adolfo A. Ferrando, James R. Downing, Tyler Jacks, H. Robert Horvitz & Todd R. Nature **435**(7043), 834 (2005).
[60] **Fei Sun**, Xia Huo, Yujia Zhai, Aojin Wang, Jianxing Xu, Dan Su, Mark Bartlam and Zihe Rao, Cell **121**, 1043-1057 (2005).
[61] **Hui Jiang**, Wei Guo, Xinhua Liang, and Yi Rao, Cell **120**,123-135 (2005).
[62] **Ji-Song Guan**, Zhen-Zhong Xu, Hua Gao, Shao-Qiu He, Guo-Qiang Ma, TaoSun, Li-hua Wang, Zhen-Ning Zhang, Isabella Lena, Ian Kitchen, Robert Elde, Andreas Zimmer, Cheng He, Gang Pel, Lan Bao, and Xu Zhang*, Cell **122**, 618-631 (2005).